\documentclass[fleqn,twoside]{article}
\usepackage{espcrc2}
\usepackage{epsfig}
\usepackage{amsmath}
\usepackage{amssymb}
\usepackage{amsbsy}
\usepackage{amsfonts}
\usepackage{graphics}
\usepackage{latexsym}
\usepackage{epsf}
\usepackage{graphicx}

\newcommand{\bc}{\begin{center}}
\newcommand{\ec}{\end{center}}
\newcommand{\be}{\begin{equation}}
\newcommand{\ee}{\end{equation}}
\newcommand{\bea}{\begin{eqnarray}}
\newcommand{\eea}{\end{eqnarray}}
\newcommand{\bi}{\begin{itemize}}
\newcommand{\ei}{\end{itemize}}
\newcommand{\bt}{\begin{tabular}}
\newcommand{\et}{\end{tabular}}

\def\Nf{N_{\rm f}}

\def\mq{m_{\rm q}}
\def\mps{m_{\rm PS}}

\def\msbar{\overline{\rm MS}}

\def\ks{\kappa_{\rm sea}}
\def\kv{\kappa_{\rm val}}

\newcommand{\Dlr}{\stackrel{\leftrightarrow}{D}}

\title{
\begin{flushleft}
\vspace*{-2cm}
{\normalsize DESY 04-189}\\[-0.3em]
{\normalsize LU-ITP 2004/029}\\[-0.3em]
{\normalsize Edinburgh 2004/21}
\vspace*{0.15cm}
\end{flushleft}
Generalized parton distributions and structure functions from 
full lattice QCD%
\thanks{Talks presented by D. Pleiter and J. Zanotti}
}

\author{M. G\"ockeler\address{Institut f\"ur Theoretische Physik, 
Universit\"at Leipzig, D-04109 Leipzig, Germany}$^,$\address{
Institut f\"ur Theoretische Physik, Universit\"at Regensburg, 
D-93040 Regensburg, Germany},
Ph. H\"agler\address{Department of Physics and Astronomy, Vrije Universiteit,
1081 HV Amsterdam, NL},
R. Horsley\address{School of Physics, University of Edinburgh, 
Edinburgh EH9 3JZ, UK},
D. Pleiter\address{John von Neumann-Institut f\"ur Computing 
NIC / DESY, 15738 Zeuthen, Germany},
P.E.L. Rakow\address{Theoretical Physics Division, Dep.~of 
Math.~Sciences, University of Liverpool, Liverpool L69 3BX, UK},
A. Sch\"afer$^{\rm b}$,
G. Schierholz$^{\rm e,}$\address{Deutsches Elektronen-Synchrotron  
DESY, 22603 Hamburg, Germany} and
J.M. Zanotti$^{\rm e}$
\emph{(QCDSF Collaboration)}}

\begin{document}

\begin{abstract}
We present here the latest results from the QCDSF
collaboration for (moments of) structure functions and generalized form
factors in full QCD with $\Nf=2$ ${\cal O}(a)$-improved Wilson fermions
based on simulations closer to the chiral and continuum limit.
\end{abstract}

\maketitle

\vspace*{-15mm}
\section{INTRODUCTION}
\vspace*{-2mm}

Understanding the internal structure of hadrons in terms of quarks and
gluons (partons), and in particular how these provide the
binding and spin of the nucleon, is one of the outstanding problems in
particle physics.

Matrix elements of the light cone operator
\vspace*{-1mm}
\begin{align}
{\cal O}(x) = \int \frac{d\lambda}{4\pi} \;&
  e^{i\lambda x}\;\overline{q}
  (-\frac{\lambda}{2} n) \not\! n \\[-2mm] &
  {\cal P}e^{\!-ig
  \int^{\lambda/2}_{-\lambda/2} d\alpha\,n\cdot A(\alpha n)}
  q(\frac{\lambda}{2} n) \nonumber
\end{align}
measured in deep-inelastic scattering experiments provide a wealth of
information regarding the quark and gluon content of the nucleon.

Expanding ${\cal O}(x)$ in terms of local operators via the operator
product expansion generates the tower of twist-2 operators
\vspace*{-2mm}
\be
{\cal O}^{\{\mu_1\cdots\mu_n\}} = \overline{q} \, i \,
\gamma^{\{\mu_1}\, \Dlr^{\mu_2} \cdots
\Dlr^{\mu_n\}}\! q\ ,
\label{twist2}
\ee
where $\Dlr = \frac{1}{2}(\overrightarrow{D} - \overleftarrow{D})$ and
$\{\cdots\}$ indicates symmetrization of indices and removal of
traces.
The (non-)forward matrix elements of Eq.~(\ref{twist2}) specify the
$(n-1)^{th}$ moments of the (generalized) parton distributions.

Parton distributions measure the probability $|\psi(x)|^2$ of finding
a parton with fractional longitudinal momentum $x$ in the fast moving
nucleon at a given transverse resolution $1/Q$. 
Generalized parton distributions (GPDs)~\cite{GPD,Ji} describe the
coherence of two different hadron wave functions
$\psi^\dagger(x+\xi/2)\,\psi(x-\xi/2)$, one where the parton carries
fractional momentum $x+\xi/2$ and one where this fraction is
$x-\xi/2$, from which further information on the transverse
distribution of partons can be drawn~\cite{Diehl,Bu}.
In the limit where the momentum transfer $\Delta$ to the nucleon is
purely transverse, i.e. $\Delta=(0,\vec{\Delta}_\perp)$ and
$\xi=0$, GPDs regain a probabilistic interpretation~\cite{Bu}.
In the forward limit ($\Delta = 0$), these distributions reduce to the
Feynman parton distributions.

Moments of GPDs are amenable to lattice
calculations~\cite{QCDSF-1,MIT,MIT-2}.  
Thus, they offer a promising way to link phenomenological observations
to first principle theoretical considerations.
In this talk we report on recent unquenched results obtained by
the QCDSF collaboration. 

\vspace*{-2mm}
\section{SIMULATION DETAILS}

We simulate with $\Nf=2$ dynamical configurations generated with
Wilson glue and non-perturbatively ${\cal O}(a)$ improved Wilson
fermions.
For five different values $\beta=5.20$, $5.25$, $5.26$,
$5.29$, $5.40$ and up to three different kappa values per beta
we have in collaboration with UKQCD generated ${\cal O}(2000-8000)$
trajectories.
Lattice spacings and spatial volumes vary between 0.075-0.123~fm and
(1.5-2.2~fm)$^3$ respectively.

Correlation functions are calculated on configurations
taken at a distance of 5-10 trajectories using 8-4 different locations of
the fermion source. We use binning to obtain an effective distance of
20 trajectories. The size of the bins has little effect on the
error, which indicates auto-correlations are small.
This work improves on previous calculations by adding one more sink
momentum, $\vec{p}_2$, and polarization, $\Gamma_1$. We use
$\vec{p}_0 = ( 0, 0, 0 )$,
$\vec{p}_1 = ( p, 0, 0 )$,
$\vec{p}_2  = ( 0, p, 0 )$
($p=2\pi/L_S$) and
$\Gamma_{\rm unpol} = \frac{1}{2}(1+\gamma_4)$,
$\Gamma_1 = \frac{1}{2}(1+\gamma_4)\, i\gamma_5\gamma_1$,
$\Gamma_2 = \frac{1}{2}(1+\gamma_4)\, i\gamma_5\gamma_2$.

\begin{figure}[t]
\bc
\vspace*{1mm}
\includegraphics[width=75mm]{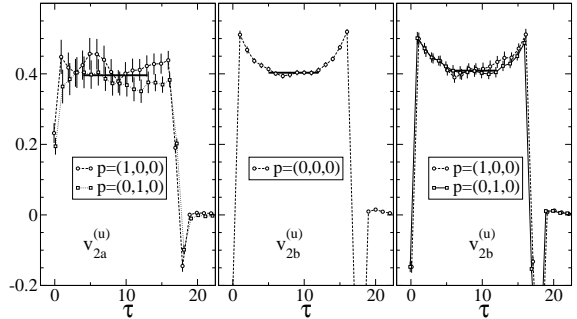}
\vspace*{-15mm}
\caption{\emph{Bare results for the lowest moment of $F_2$ at
$(\beta,\ks) = (5.4,0.13560)$, $t=17$.
The fat horizontal lines are obtained from a fit to the data.
$v_{\rm 2a}$ and $v_{\rm 2b}$ refer to different lattice
versions of the same operator.}}
\label{signal}
\ec
\vspace*{-10mm}
\end{figure}

\begin{figure}[b]
\bc
\vspace*{-5mm}
\includegraphics[width=75mm]{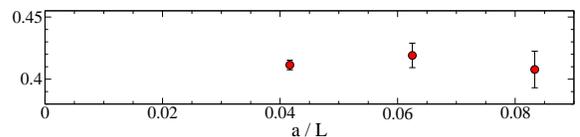}
\vspace*{-14mm}
\caption{\emph{Finite size analysis for $v_{2b}^{\rm RGI}$ ($\vec{p} =
  (0,0,0)$) at $(\beta,\ks)=(5.29,0.13550)$ on 
  $V=24^3\times 48$, $16^3\times 32$, $12^3\times 32$ lattices.}}
\label{v2b-fse}
\ec
\vspace*{-0.5cm}
\end{figure}

\begin{figure}[t]
\bc
\vspace*{1mm}
\includegraphics[width=75mm]{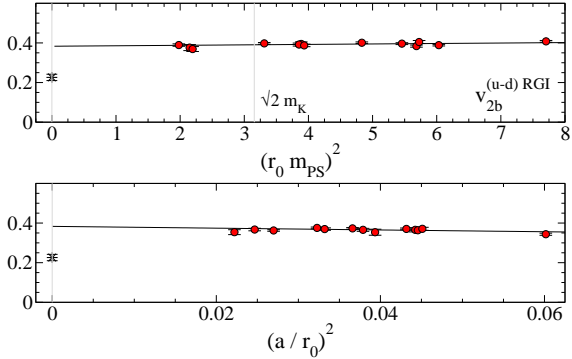}
\vspace*{-14mm}
\caption{\emph{%
Results from a fit to Eq.~(\ref{chiralextrapolation}) with
$\Lambda_{\chi} = 0$. The upper plot shows the chiral extrapolation
$v_{2b}^{\rm RGI}(r_0,\mps) - c_{2b} (a/r_0)^2$
and the lower plot the continuum extrapolation
$v_{2b}^{\rm RGI}(r_0,\mps) - F^{v_{2b}}(r_0 \mps)$.
}}
\label{fig:v2b-lin}
\ec
\vspace*{-13mm}
\end{figure}

\section{QUARK DISTRIBUTIONS}

The moments of the $F_2$ structure function $\langle x^{n-1}\rangle$
are determined by calculating the matrix elements
\begin{multline}
\langle N(\vec{p}) |
\left[ {\cal O}^{\{\mu_1\cdots\mu_n\}}_q - {\rm Tr} \right] |
N(\vec{p}) \rangle^{\cal S} = \\
2 v_n^{(q){\cal S}}(g^{\cal S}({\cal M}))\;
[p^{\mu_1}\cdots p^{\mu_n} - {\rm Tr}],
\end{multline}
and corresponding Wilson coefficients
in a renormalization scheme $\cal S$ (eg.~$\msbar$) and at a scale
$\cal M$ (e.g.~2 GeV).
See \cite{QCDSF-timid} for the operators to use on the lattice.
To eliminate ${\cal O}(a)$ terms in the matrix elements, these operators
need to be improved. For the lowest moment, i.e.~$v_2$, this amounts
to replacing ${\cal O}^{\gamma}_{\mu\nu}$ by
$(1 + a \mq c_0)\;{\cal O}^{\gamma}_{\mu\nu} +
\sum_{i=1}^2 a\,c_i\;{\cal O}^{(i)}_{\mu\nu}$.
The improvement coefficient $c_0$ is only known perturbatively, while
the other coefficients are unknown. Since the corresponding operator
matrix elements turn out to be negligible we will set $c_1 = c_2 =
0$.
For the higher moments an additional problem arises: the operators may
mix with operators with total derivatives \cite{QCDSF-renorm}.
However, the corresponding operators are again found to be small and
are therefore neglected here.

Matrix elements are determined from the ratio of three-point to
two-point correlation functions
\bea
\label{eq:ratio}
\lefteqn{{\cal R}(t,\tau;\vec{p}\,',\vec{p};{\cal O})\, =
 \frac{C_\Gamma (t,\tau;\vec{p}\,', \vec{p},{\cal O})} 
        {C_2(t,\vec{p}\,')}} \\\nonumber
&&\times \left[
  \frac{C_2(\tau,\vec{p}\,') C_2(t,\vec{p}\,') C_2(t-\tau,\vec{p}\,)}
  {C_2(\tau,\vec{p}\,) C_2(t,\vec{p}\,) C_2(t-\tau,\vec{p}\,')}
\right]^{\frac{1}{2} }
\eea
where $C_2$ is the unpolarized baryon two-point function with a source
at time 0 and sink at time $t$, while the unpolarized three-point
function $C_\Gamma$ has an operator ${\cal O}$ insertion at time
$\tau$.  For the matrix elements $v_n$ we have $\vec{p} = \vec{p}\,'$,
i.e.~the last term of Eq.~(\ref{eq:ratio}) is equal to 1.
To improve our signal for non-zero momentum we average over the
results for $C_\Gamma$ and $C_2$ for both momenta before we calculate
the ratio $\cal R$.
In Fig.~\ref{signal} we compare the results for the different ways
of obtaining the lowest moment. Within errors the results are consistent.

The bare matrix elements must be renormalized. While non-perturbatively
determined renormalization factors are known in
quenched QCD \cite{NP}, the situation for dynamical
simulations is still under investigation.
Here we use tadpole-improved renormalization-group-improved boosted
perturbation theory \cite{TRB-PT} to convert our lattice results
to RGI.

Before we examine the quark mass and lattice spacing behaviour of our
results, we first check for finite size effects.
In Fig.~\ref{v2b-fse} we present the first results of a finite size
analysis for $v_{2b}$. Here we plot $v_{2b}$ as a function of inverse
spatial lattice extent for three different volumes $V=24^3\times 48$,
$16^3\times 32$, $12^3\times 32$ at $(\beta,\ks)=(5.29,0.13550)$.
This preliminary analysis reveals that the finite size effects for
$v_{2b}$ are small.

Finally, the discretization effects and quark mass dependence need
to be investigated. The commonly accepted procedure is to first extrapolate
to the continuum limit then to the chiral limit.
Since the currently available data does not allow us to perform both
extrapolations separately, we make the following ansatz for
a simultaneous chiral and continuum extrapolation:
\begin{multline}
v_n^{\rm RGI}(r_0,\mps) = 
  F^{v_n}(r_0\mps)\; + \\
  c_n\,\left(\frac{a}{r_0}\right)^2 +
  d_n\,a r_0\,\mps^2.
\label{extrapolation}
\end{multline}
The terms proportional to $c_n$ and $d_n$ take discretization errors
$\propto a^2$ and residual effects $\propto a m_q$ into account
(we fix $d_n = 0$).
The first term corresponds to a chiral extrapolation.
To incorporate chiral physics into the extrapolation taking into
account the effects due to the `pion cloud' surrounding the nucleon,
the ansatz
\be
F^{v_n}(x) =
  v_n^{\rm RGI} 
  \left( 1 - c x^2 \ln \frac{x^2}{x^2 + r_0^2 \Lambda_{\chi}^2}\right) +
  a_n x^2
\label{chiralextrapolation}
\ee
has been proposed \cite{Detmold},
where $c\approx 0.663$ and $\Lambda_\chi$ is a free parameter,
usually taken to be of ${\cal O}(1\, {\rm GeV})$.
Setting $\Lambda_\chi = 0$ reduces
Eq.~(\ref{chiralextrapolation}) to a linear extrapolation in
$(r_0 \mps)^2$. 

\begin{figure}[t]
\bc
\vspace*{1mm}
\includegraphics[width=75mm]{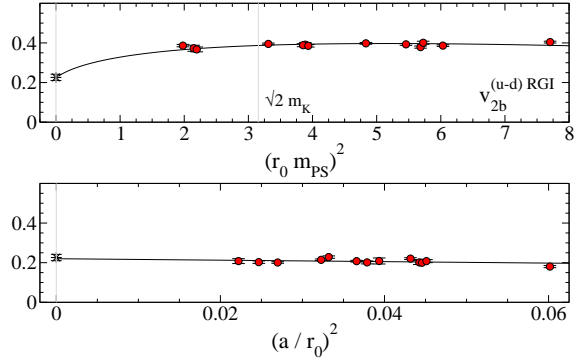}
\vspace*{-12mm}
\caption{\emph{Same as Fig.~\ref{fig:v2b-lin} but for $\Lambda_\chi=1$~GeV.}}
\label{fig:v2b-chiral}
\ec
\vspace*{-10mm}
\end{figure}

\begin{figure}[t]
\bc
\vspace*{0mm}
\includegraphics[height=75mm,angle=-90]{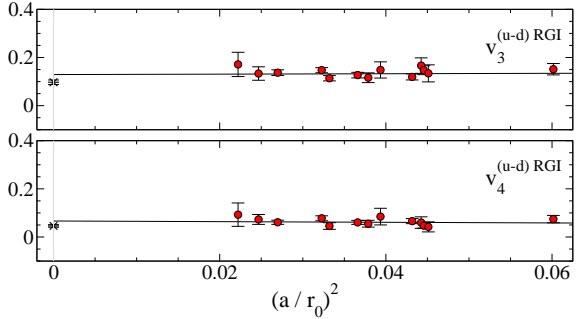}
\vspace*{-11mm}
\caption{\emph{%
Continuum extrapolation of $v_3$ and $v_4$.
}}
\label{fig:v34}
\ec
\vspace*{-10mm}
\end{figure}

In Fig.~\ref{fig:v2b-lin}, we plot the results for $v_{2b}^{\rm
  RGI}(r_0,\mps) - c_{2b} (a/r_0)^2$ as calculated from
Eq.~(\ref{extrapolation}) with $\Lambda_\chi = 0$
($\chi^2/{\rm d.o.f} = 2$)
and plotted as a function of $\mps^2$ for the isovector $u-d$.
We see here that although our data at heavy quark masses agree very
well with a linear extrapolation, the predicted value in the chiral
limit is roughly twice the experimental value.
A fit using Eq.~(\ref{extrapolation}) with $\Lambda_\chi = 1$~GeV
and the result in the chiral limit constrained
to the experimental value does not describe our data equally well
($\chi^2/{\rm d.o.f} = 5$).
However, as can be seen from Fig.~\ref{fig:v2b-chiral} any
disagreement is not significant since all of the curvature of the fit
occurs in the light quark mass region where there is no data.
Even at our lightest quark mass ($\mps\approx 500$MeV) we are not yet
into the region where curvature is expected to occur.

Turning our attention now to discretization effects, in the bottom
figure of Fig.~\ref{fig:v2b-chiral} we plot $v_{2b}^{\rm RGI}(r_0,\mps) -
F^{v_{2b}}(r_0\mps)$ as a function of $(a/r_0)^2$. Here we observe a
very small dependence on the lattice spacing, indicating that not only
is our ${\cal O}(a)$ improvement program working, but also that
${\cal O}(a^2)$ effects are small.

For the higher moments we only perform a chiral extrapolation linear
in $(r_0 \mps)^2$. The results for $v_3$ and $v_4$ from the combined
chiral and continuum extrapolation are shown in Fig.~\ref{fig:v34}.

\begin{figure}[t]
\vspace*{1mm}
\bc
\includegraphics[height=7.5cm,angle=90]{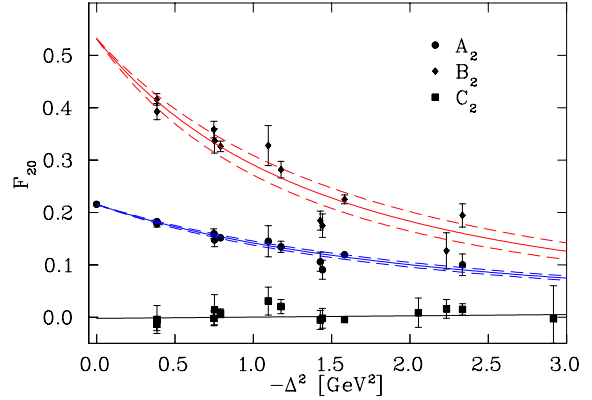}
\vspace*{-12mm}
\caption{\emph{Generalized form factors $A^{u-d}_{20},\ B^{u-d}_{20}$ and
  $C^{u-d}_{2}$, together with a dipole fit.}}
\label{ABC20}
\ec
\vspace*{-10mm}
\end{figure}

\section{GENERALIZED PARTON DISTRIBUTIONS}

Non-forward matrix elements of the twist-2 operators in
Eq.~(\ref{twist2}) yield the moments of generalized parton
distributions 
\bea
\!\!\int_{-1}^1 dx\, x^{n-1}\, H_q(x,\xi, \Delta^2) \!\!&=&\!\! 
 H_{qn}(\xi,\Delta^2) \nonumber \\
\!\!\int_{-1}^1 dx\, x^{n-1}\, E_q(x,\xi, \Delta^2) \!\!&=&\!\! 
 E_{qn}(\xi,\Delta^2) 
\eea
where \cite{Ji}
\bea
H_{qn}(\xi,\Delta^2) &=& \sum_{i=0}^{\frac{n-1}{2}}
 A_{qn,2i}(\Delta^2) (-2\xi)^{2i} \\
 &+& {\rm Mod}(n+1,2) C_{qn}(\Delta^2) (-2\xi)^n \nonumber\\
E_{qn}(\xi,\Delta^2) &=& \sum_{i=0}^{\frac{n-1}{2}}
 B_{qn,2i}(\Delta^2) (-2\xi)^{2i} \nonumber\\
 &-& {\rm Mod}(n+1,2) C_{qn}(\Delta^2) (-2\xi)^n \nonumber 
\label{GPDmoments}
\eea
and the generalized form factors $A^q_{n,2i}(\Delta^2)$,
$B^q_{n,2i}(\Delta^2)$ and $C^q_{n}(\Delta^2)$ for the lowest three moments
are extracted from the nucleon matrix elements $\langle p'|{\cal
  O}^{\{\mu_1\cdots\mu_n\}}|p\rangle$ \cite{Ji}.
For the lowest moment, $A_{10}$ and $B_{10}$ are just the Dirac and
Pauli form factors $F_1$ and $F_2$, respectively. We also observe that
in the forward limit ($\Delta^2 = \xi = 0$), the moments of $H_q$
reduce to the moments of the unpolarized parton distribution $A_{n0}
= \langle x^{n-1}\rangle$.

In order to extract the non-forward matrix elements 
we compute ratios of three- and two-point function as in
Eq.~(\ref{eq:ratio}) with $\vec{p}\ne \vec{p}\,'$.

In Fig.~\ref{ABC20} we show, as an example, the generalized form
factors $A_{20},\ B_{20}$ and $C_{2}$ for the non-singlet, $u-d$, on
a $24^3\times48$ lattice at $\beta=5.40$ and $\ks=\kv=0.13500$
corresponding to a lattice spacing, $a\,r_0 = 6.088$ and
$\mps\approx970$ MeV.

The generalized form factors $A_{20},\ B_{20}$ are well described
by the dipole ansatz
\be
A_n^q (\Delta^2) = \frac{A_n^q(0)}{\big( 1 - {\Delta^2/M_n^2}
  \big)^2} \ , 
\label{dipole}
\ee
while $C^{u-d}_{2}$ is consistent with zero.

\begin{figure}[t]
\bc
\includegraphics[height=7.5cm,angle=90]{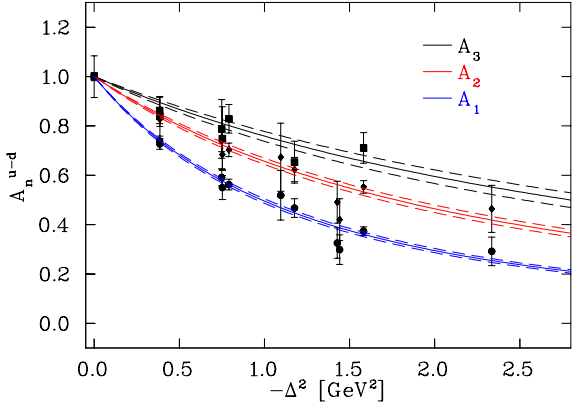}
\vspace*{-12mm}
\caption{\emph{Generalized form factors $A^{u-d}_{10},\ A^{u-d}_{20}$ and
  $A^{u-d}_{30}$, together with a dipole fit. All form factors have
  been normalized to unity.}}
\label{A123}
\ec
\vspace*{-12mm}
\end{figure}

Burkardt~\cite{Bu} has shown that the spin-independent and
spin-dependent generalized parton distributions $H(x,0,\Delta^2)$ and
$\widetilde{H}(x,0,\Delta^2)$ gain a physical interpretation when
Fourier transformed to impact parameter space
\be
q(x,\vec{b}_\perp) =  \int \frac{d^2\Delta_\perp}{(2\pi)^2}\,
{\rm e}^{-i\vec{b}_\perp \cdot \vec{\Delta}_\perp}
H(x,0,-\Delta_\perp^2)\, ,
\label{fourier}
\ee
(and similar for the polarized $\Delta q(x,\vec{b}_\perp)$) where
$q(x,\vec{b}_\perp)$ is the probability of finding a quark with
longitudinal momentum fraction $x$ and at transverse position (or
impact parameter) $\vec{b}_\perp$.

Burkardt \cite{Bu} also argued that $H(x,0,-\Delta_\perp^2)$ becomes
$\Delta_\perp^2$-independent as $x\rightarrow 1$ since, physically, we
expect the transverse size of the nucleon to decrease as $x$
increases,
i.e. $\lim_{x\rightarrow 1} q(x,\vec{b}_\perp) \propto
\delta^2(\vec{b}_\perp)$. As a result, we expect the slopes of the
moments of $H(x,0,-\Delta_\perp^2)$ in $\Delta_\perp^2$ to decrease as
we proceed to higher moments. 
This is also true for the polarized moments of
$\widetilde{H}(x,0,-\Delta_\perp^2)$, so from Eq.~(\ref{GPDmoments})
with $\xi=0$, we expect that the slopes of the generalized form
factors $A_{n0}(\Delta^2)$ and $\widetilde{A}_{n0}(\Delta^2)$ should
decrease with increasing $n$.

\begin{figure}[t]
\bc
\includegraphics[height=7.5cm,angle=90]{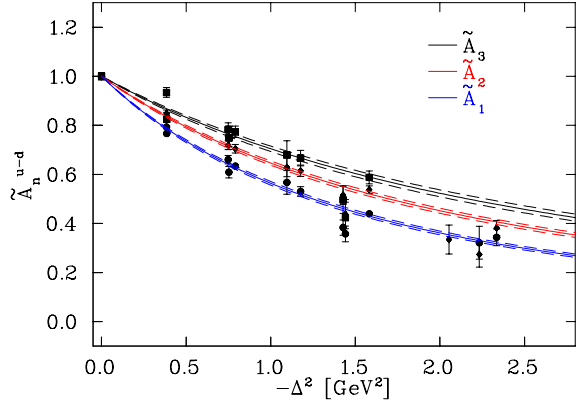}
\vspace*{-12mm}
\caption{\emph{Same as Fig.~\protect{\ref{A123}} but for
  $\tilde{A}^{u-d}_{n0}$.}}
\label{Apol123}
\ec
\vspace*{-12mm}
\end{figure}

In Figs.~\ref{A123} and \ref{Apol123}, we show the
$\Delta^2$-dependence of $A_{n0}(\Delta^2)$ and
$\widetilde{A}_{n0}(\Delta^2)$, respectively for $n=1,2,3$.
The form factors have been normalized to unity to make a comparison of
the slopes easier and as in Fig~\ref{ABC20} we fit the form
factors with a dipole form as in Eq.~(\ref{dipole}).
We note here that the form factors for the
unpolarized moments are well separated and that their
slopes do indeed decrease with increasing $n$ as predicted.
For the polarized moments, we observe a similar scenario, however here
the change in slope between the form factors is not as large. This is
to be compared with the results from Ref.~\cite{MIT-2} which reveal no
change in slope between the $n=2$ and $n=3$ polarized moments.

\begin{figure}[t]
\bc
\includegraphics[height=7.5cm,angle=90]{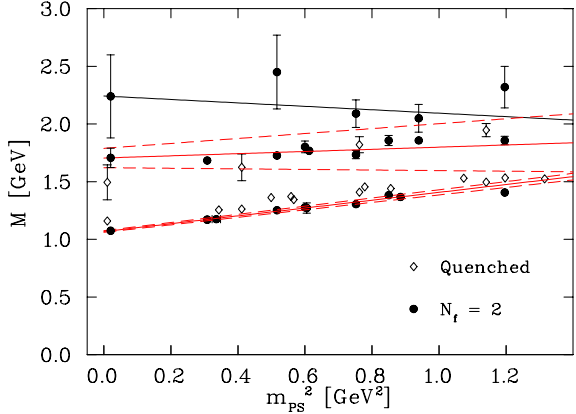}
\vspace*{-12mm}
\caption{\emph{Dipole masses for the first three moments of
  $H(x,0,-\Delta_\perp^2)$ as a function of $\mps^2$, together with a
linear extrapolation to the physical pion mass.}}
\label{dipolemasses}
\ec
\vspace*{-10mm}
\end{figure}

Although fitting the form factors with a dipole is purely
phenomenological, it does provide us with a useful means to measure the
change in slope of the form factors by monitoring the extracted dipole
masses $(M_1,\, M_2,\, M_3)$ as we proceed to higher moments.
We have calculated these generalized form factors on a subset of our
full complement of $(\beta,\, \kappa)$ combinations and have extracted
the corresponding dipole masses. We plot these dipole masses in
Fig.~\ref{dipolemasses} as a function of $\mps^2$. 
Included for comparison are previous quenched results for the first 
\cite{QCDSF} and second \cite{QCDSF-1} moments. The new unquenched
results indicate that quenching effects are small for form factors.

The important feature to note in Fig.~\ref{dipolemasses} is the
distinct separation between (and increase in magnitude of) the dipole
masses as we move to higher moments $(x\rightarrow 1)$.
Although the data available for $M_3$ is limited, the behaviour of all
dipole masses appears to be linear with $\mps^2$.
Consequently, we perform individual linear extrapolations of the
dipole masses $M_1,\, M_2,\, M_3$ to the physical pion mass, although
the findings of Ref.~\cite{Thomas} suggest that the chiral
extrapolation of the dipole masses of the electromagnetic form factors
may be non-linear.

If the dipole behavior Eq.~(\ref{dipolemasses}) continues to hold for
the higher moments as well, and if we assume that the dipole masses
continue to grow in a Regge-like fashion, we may write
\be
\int_{-1}^1 \mbox{d}x\, x^{n-1} \,H_q(x,0,\Delta^2) = 
\frac{\langle x_q^{n-1}\rangle}{(1-\Delta^2/M_{n}^2)^2},
\label{H}
\ee
with $M_l^2 = \mbox{const.} + l/\alpha'$, where $\mbox{const.} \approx
-0.5$ GeV$^2$ and $1/\alpha' \approx 1.1$ GeV$^2$. 
This is sufficient to compute $H_q(x,0,\Delta^2)$ by means of an
inverse Mellin transform~\cite{DIS}.

Having done so, the desired probability distribution of finding a
parton of momentum fraction $x$ at the impact parameter
$\vec{b}_\perp$ can then be obtained by the Fourier transform of
Eq.~(\ref{fourier}).

\vspace*{-3mm}
\section{CONCLUSION}
\vspace*{-2mm}

We presented an update on our results for $\langle x^{n-1}\rangle$ in full
QCD. While data closer to the chiral and continuum limit became available,
the linear behaviour observed previously \cite{QCDSF-Cairns} still holds.
Finally, we have shown preliminary results for the first 3 moments of
the GPDs. Our data confirms the
expected flattening going to higher moments.

\vspace*{-3mm}
\section*{ACKNOWLEDGEMENTS}
\vspace*{-2mm}

The numerical calculations have been performed on the Hitachi SR8000 at
LRZ (Munich), on the Cray T3E at EPCC (Edinburgh) under PPARC
grant PPA/G/S/1998/00777 \cite{UKQCD}, and on the APEmille at NIC/DESY
(Zeuthen). This work is supported in part by DFG and the EC under contract 
HPRN-CT-2000-00145. We thank A.~Irving for providing $r_0/a$ prior to
publication.

\end{document}